\documentclass[runningheads]{llncs}
\usepackage{graphicx}
\usepackage{amssymb, amsmath}
\usepackage{verbatim, float}
\usepackage{mathrsfs}
\usepackage[usenames,dvipsnames]{pstricks}
\usepackage{epsfig}
\usepackage{pst-grad} 
\usepackage{pst-plot} 
\usepackage{cases}
\usepackage{bm}
\usepackage{multicol, algorithm}

\newcommand{\A}{{\mathcal A}}

\newcommand{\bM}{{\boldsymbol{M}}}

\newcommand{\bQ}{{\boldsymbol Q}}

\newcommand{\bV}{\boldsymbol{V}}
\newcommand{\ba}{{\boldsymbol a}}

\newcommand{\bh}{{\boldsymbol h}}

\newcommand{\bee}{{\boldsymbol{e}}}
\newcommand{\bc}{{\boldsymbol{c}}}

\newcommand{\bx}{{\boldsymbol{x}}}
\newcommand{\bz}{{\boldsymbol{z}}}

\newcommand{\bbF}{{\mathbb F}}
\newcommand{\bbN}{{\mathbb N}}

\newcommand{\bbZ}{{\mathbb Z}}

\newcommand{\ff}{\mathbb{F}}
\newcommand{\fn}{\mathbb{F}^n}
\newcommand{\fl}{\mathbb{F}^{\ell}}

\newcommand{\fp}{\mathbb{F}_p}

\newcommand{\fmn}{\mathbb{F}^{m\times n}}
\newcommand{\bbG}{\mathbb{G}}

\newcommand{\fqmn}{\mathbb{F}_q^{m\times n}}

\newcommand{\al}{\alpha}

\newcommand{\vep}{\varepsilon}

\newcommand{\define}{\stackrel{\mbox{\tiny $\triangle$}}{=}}

\newcommand{\xon}{(x_1,x_2,\ldots,x_n)}
\newcommand{\cbx}{C(\bm{x})}
\newcommand{\wax}{w_{\bm{a}}(\bx)}

\newcommand{\pp}{{\sf{pp}}}
\newcommand{\sk}{{\sf{sk}}}
\newcommand{\aux}{{\sf{aux}}}

\newcommand{\setup}{{\sf{Setup}}}
\newcommand{\comgen}{{\sf{CommitmentGen}}}
\newcommand{\quegen}{{\sf{QueriesGen}}}
\newcommand{\ansgen}{{\sf{AnswersGen}}}
\newcommand{\witgen}{{\sf{WitnessGen}}}
\newcommand{\extract}{{\sf{Extract}}}
\newcommand{\verify}{{\sf{Verify}}}
\newcommand{\bggen}{{\sf{BGGen}}}
\newcommand{\ext}{{\sf{Extract}}}
\newcommand{\pr}{{\sf{Pr}}}
\newcommand{\bg}{{\sf{BG}}}
\newcommand{\negl}{{\sf{negl}}}
\newcommand{\poly}{{\sf{poly}}}
\newcommand{\EXP}{{\sf{EXP}}_{\A,\Pi}(k, m,n,\bx,i,B)}
\newcommand{\lmc}{\text{LMC}}
\newcommand{\pir}{\text{PIR}}

\newcommand{\haj}{\hat{a}_j}
\newcommand{\hwj}{\hat{w}_j}
\newcommand{\xhi}{\hat{x}_i}

\newcommand{\hhi}{\hat{h}_i}
\newcommand{\alj}{\alpha^j}

\newcommand{\zp}{\mathbb{Z}_p}
\newcommand{\vo}{v^{(1)}}

\newcommand{\vs}{v^{(s)}}
\newcommand{\vt}{v^{(t)}}

\newcommand{\kn}{{$k$-server $n$-dimensional}{ }}
\newcommand{\kmn}{{$k$-server $m\times n$-dimensional}{ }}
\newcommand{\rand}{{ }{\leftarrow}\vcenter{\hbox{\tiny\rmfamily\upshape\$}}{ }}

\begin{document}
\title{Committed Private Information Retrieval\thanks{\scriptsize {Supported by the Australian Research Council through the Discovery Project under Grant DP200100731. The work of Hong Yen Tran was partly done when she was with RMIT University.}}}

\author{Quang Cao\inst{1}\orcidID{0000-0001-9649-943X} 
\and Hong Yen Tran\inst{2} \and Son Hoang Dau\inst{1} \and Xun Yi\inst{1}  \and Emanuele Viterbo\inst{3} \and Chen Feng\inst{4} \and Yu-Chih Huang\inst{5} \and Jingge Zhu\inst{6} \and Stanislav Kruglik\inst{7} \and Han Mao Kiah\inst{7}}
\authorrunning{C. Quang et al.}
%
\institute{RMIT University, Melbourne, Australia \\
\email{\{nhat.quang.cao2, sonhoang.dau, xun.yi\}@rmit.edu.au}\\
\and The University of New South Wales, Canberra, Australia
\email{hongyen.tran@unsw.edu.au}\\
\and Monash University, Melbourne, Australia\\
\email{emanuele.viterbo@monash.edu}\\
\and The University of British Columbia, Kelowna, Canada\\
\email{chen.feng@ubc.ca}\\
\and NYCU University, Hsinchu, Taiwan\\
\email{jerryhuang@nctu.edu.tw}\\
\and The University of Melbourne, Melbourne, Australia\\
\email{jingge.zhu@unimelb.edu.au}\\
\and Nanyang Technological University, Singapore\\
\email{\{stanislav.kruglik, hmkiah\}@ntu.edu.sg}
}

\maketitle              
\vspace{-10pt}

\begin{abstract}
A \textit{private information retrieval} (PIR) scheme allows a client to retrieve a data item $x_i$ among $n$ items $x_1,x_2,\ldots,x_n$ from $k$ servers, without revealing what $i$ is even when $t < k$ servers collude and try to learn $i$. Such a PIR scheme is said to be \textit{$t$-private}.
A PIR scheme is \textit{$v$-verifiable} if the client can verify the correctness of the retrieved $x_i$ even when $v \leq k$ servers collude and try to fool the client by sending manipulated data. Most of the previous works in the literature on PIR assumed that $v < k$, leaving the case of \textit{all-colluding} servers open. We propose a generic construction that combines a \textit{linear map commitment} (LMC) and an arbitrary linear PIR scheme to produce a $k$-verifiable PIR scheme, termed a \textit{committed PIR} scheme. Such a scheme guarantees that even in the worst scenario, when all servers are under the control of an attacker, although the privacy is unavoidably lost, the client won't be fooled into accepting an incorrect $x_i$. We demonstrate the practicality of our proposal by implementing the committed PIR schemes based on the Lai-Malavolta LMC and three well-known PIR schemes using the \texttt{GMP} library and \texttt{blst}, the current fastest C library for elliptic curve pairings. 
\keywords{Private information retrieval \and verifiability \and  malicious server \and commitment scheme \and pairing \and elliptic curve.}
\end{abstract}

\vspace{-20pt}
\section{Introduction}
\label{sec:intro}

In this work, we revisit private information retrieval (PIR), a classic tool in cryptography, and investigate the extent that PIR can be used in a \textit{trustless} system in which participants can be corrupted. While the basic PIR only provides privacy, i.e., making sure that a client can privately retrieve a data item of interest without revealing it to any server that stores the collection of data, we are interested in three extra \textit{security}\footnote{\scriptsize {In the PIR literature, `security' was often used to refer to the concept of `verifiability' defined in this work. However, in our opinion, `security' is a rather broad term and should not be used as the name of a specific property. We make an effort to fix that terminology issue in this work, using `security' as an umbrella term instead.}} requirements, namely, \textit{verifiability}, \textit{accountability}, and \textit{Byzantine-robustness}. These requirements are made under the assumption that a group of malicious servers are not only keen on learning the retrieved data but also on making the client recover \textit{incorrect} data to achieve certain purposes. This type of security requirements are crucial to extend the usage of PIR beyond trusted systems, which make sense mostly in theory, to the more practical \textit{trustless} systems, which are capable of governing both trusted and malicious parties. Note that `trusted' is also a very shaky status: even a supposedly trusted party like a well-established bank or a government agency can still be attacked and temporarily become a malicious party, which 
may cause severe damage to the customers (see, e.g. devastating attacks on Australian universities, Medibank, Optus, and Fire Rescue Victoria in 2022~\cite{UribeAlice2022CoOP,biddle2022public}).

A basic \textit{private information retrieval} (PIR) scheme allows a client to download a data item $x_i$ among a collection of $n$ items $x_1,x_2,\ldots,x_n$ from $k\geq 1$ servers without revealing the index $i$ to any curious server.  
The very first private information retrieval (PIR) scheme with two servers was introduced in the seminal work of Chor-Kushilevitz-Goldreich-Sudan~\cite{Chor-Kushilevitz-Goldreich-Sudan_JACM1998}, which works as follows.
The two servers both store $x_1,x_2,\ldots,x_n$, which are elements from a finite field $\bbF$ of characteristic~2.
To privately retrieve $x_i$, the client selects a random set $J\subseteq \{1,2,\ldots,n\}$ and requests $\sum_{j \in J}x_j$ from Server 1 and $x_i + \big(\sum_{j \in J}x_j\big)$ from Server 2. As $\bbF$ has characteristic 2, the client can simply add the two answers to extract $x_i$. Moreover, as $J$ is a random set, from the query, each server achieves no information (in Shannon's sense) about $i$. 
We refer to this as the CKGS scheme and use it as a toy example to demonstrate our approach below.

The CKGS scheme, while providing privacy against an \textit{honest-but-curious} server, doesn't protect the client against a \textit{malicious} one: if the malicious server sends an incorrect answer, the client will end up with an incorrect data item $\xhi \neq x_i$. 
To construct a secure PIR scheme that can deal with malicious servers, there are two approaches: the \textit{joint-design approach} (a PIR scheme is designed with built-in security) and the \textit{modular approach} (combining a PIR and another cryptographic primitive, both of which are separately designed).
As far as we know, most related works in the literature~\cite{KeZhang_ISIT2022,ZhaoWangHuang_IS2021,ZhangWang_SP2022,DevetGoldbergHeninger_UsenixSS12,YangBennett_AICSA2002,BeimelStahl_SCN2002,BeimelStahl_JC2007,Goldberg_SP2007} (except for~\cite{ZhangSafaviNaini_ACNS2014}) followed the former.
While the first approach requires more tailor-made designs, which are harder to develop but potentially achieve better performance, the second provides greater simplicity and flexibility: an arbitrary PIR scheme and an arbitrary commitment scheme will work together to achieve a secure PIR scheme. Moreover, any improvement in either PIR or commitment schemes will automatically translate to an improvement to this approach.
In the cope of this work, we focus on the second approach, applying a \textit{commitment scheme} on top of a PIR. The gist of this approach is to publish a digest of the data, referred to as the \textit{commitment}, before the PIR session starts. Once the commitment has been produced and made public, the client can use the commitment to confirm the correctness of its desired data item, even when \textit{all} servers are malicious. 

An obvious commitment-based solution that allows the client to verify the correctness of its derived data is using (cryptographic) hashes of the data as the commitment: the hashes $h_j=h(x_j)$, $j = 1,2,\ldots,n$ are made public before the PIR session starts, and then the client can download all the hashes\footnote{\scriptsize {The client can gather the hashes by downloading them from the data owner, or from a few random servers in a decentralized system (e.g. a blockchain) and using a majority vote to determine the correct $h_i$.}} and perform a hash verification on the derived $\xhi$ and accepts it if $h(\xhi)=h_i$.
This solution, however, increases the download cost for the client due to the extra $sn$ hashes coming from $s$ servers for some constant $s$. More importantly, this makes the PIR protocol cumbersome and unsuitable to systems requiring compact data-commitments such as the blockchains, where the commitment to the data (transactions, chain states) is often a single 256-bit hash (the Merkle proof) stored in a small block header of a rather limited size, e.g. 80 bytes in Bitcoin and around 500 bytes in Ethereum. 
Here, a potential application in this context is for a client to privately retrieve a transaction in a block.

\vspace{-10pt}
\begin{figure}
    \centering
    \includegraphics[scale=0.88]{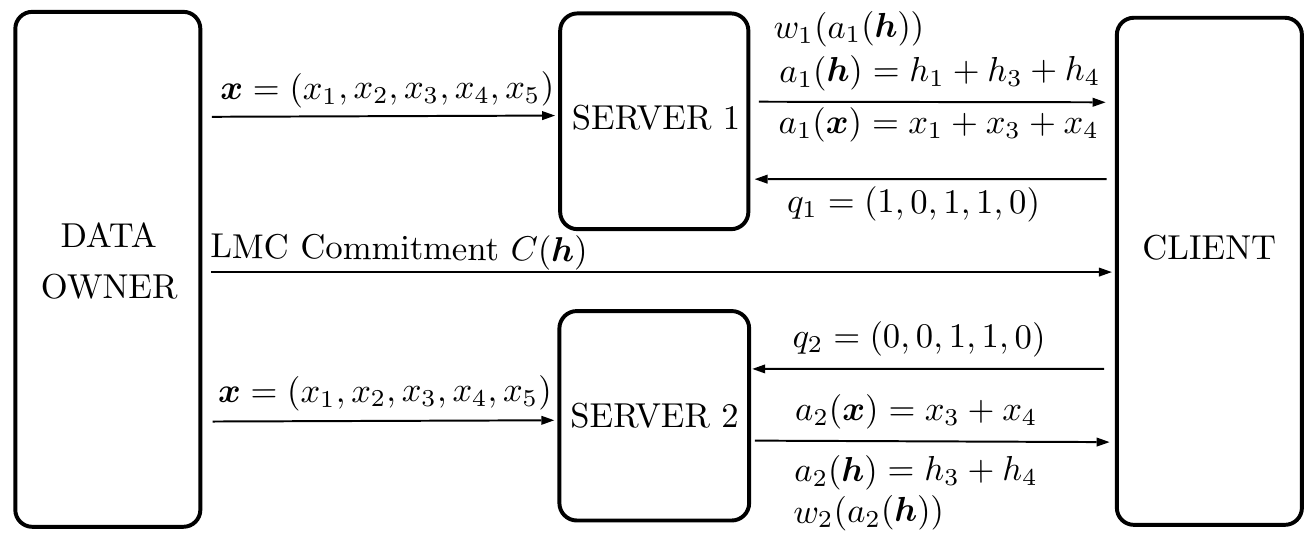}
    \vspace{-10pt}
    \caption{An example of a 2-server committed PIR scheme based on an LMC and the CKGS PIR scheme~\cite{Chor-Kushilevitz-Goldreich-Sudan_JACM1998} (see Example~\ref{ex:LMC_Chor} for more details). The client privately requests both $x_1$ and $h_1 = h(x_1)$, where the correctness of the retrieved \textit{hash} $\hat{h}_1$ can be verified thanks to the LMC. It can then verify the correctness of the retrieved \textit{data} $\hat{x}_1$ by performing a hash verification $h(\hat{x}_1) \overset{?}{=} \hat{h}_1$. The size of the $x_j$'s can be arbitrarily large. The commitment $C(\bh)$ and the witnesses $w_1(a_1(\bh)), w_2(a_2(\bh))$ are of size only 384 bits (equivalent to 1.5$\times$ SHA3-256 hash) if the Lai-Malavolta LMC~\cite{LaiMalavolta_CRYPTO2019} is used.}
    \vspace{-10pt}
    \label{fig:toy1}
\end{figure}

We address the aforementioned drawback of the hash-based solution by using linear map commitments with a \textit{constant-size} commitment on top of linear PIR schemes to provide verifiability.
A linear map commitment (LMC)~\cite{LibertRamannaYung_LIPIcs16,PeikertPepinSharp_TCC2021,LaiMalavolta_CRYPTO2019,CampanelliNitulescuRafolsZacharakisZapico_ePrint2022} allows the prover to generate a \textit{commitment} $C(\bx)$ to a vector $\bx = (x_1,x_2,\ldots,x_n)\in \bbF^n$ and a \textit{witness} $\wax$ based on which the verifier can verify that a retrieved value $y\in \bbF$ is indeed the value of the linear combination $\ba\cdot\bx=\sum_{j=1}^n a_jx_j$.
In a compact LMC, e.g.~\cite{LaiMalavolta_CRYPTO2019}, the sizes of the commitment $C(\bx)$ and the witness $\wax$ (in bits) are constant, i.e. together their sizes are equivalent to a couple of field elements only.
To make the PIR scheme suitable to a database with large-size data items, we let $x_i\in \ff^m$ 
where $m$ can be arbitrarily large and the LMC can be applied instead to the hashes $\bh = (h_1,h_2,\ldots,h_n)$ of $\bx \in \fmn$ (see Fig.~\ref{fig:toy1} for a toy example). Our proposed scheme, referred to as the \textit{committed PIR} scheme, provides $k$-verifiability: even in the extreme case where all servers are attacked and controlled by a malicious attacker, although the privacy is unavoidably lost, the scheme is still capable of protecting the client from accepting fault data.

Our main contributions are summarized below.
\begin{itemize}
    \item We propose a novel modular approach that combines a \textit{linear map commitment} scheme and a linear PIR scheme to construct a \textit{committed PIR} scheme that provides verifiability 
    on top of the traditional privacy. Our proposed scheme is capable of preventing the client from accepting an incorrect data item, even when \textit{all} $k$ servers are malicious and send manipulated data. Most previous works can only tolerate up to $k-1$ malicious servers.
    \item We carry out three case studies discussing the constructions of committed PIR schemes using a specific linear map commitment on top of the three well-known representative PIR schemes: the CKGS scheme~\cite{Chor-Kushilevitz-Goldreich-Sudan_JACM1998} (the very first PIR scheme), the WY scheme~\cite{WoodruffYekhanin_SJC2007} (lowest \textit{upload cost}, aka query size), and the BE scheme~\cite{BitarRouayheb_ITW2018} (lowest \textit{download cost}, aka answer size). The LMC primitive incurs only a constant-size communication overhead.
    \item We implemented all three schemes in C on top of the \texttt{GMP} library (for efficient handling of large numbers) and the \texttt{blst} library (the current fastest library for elliptic curve operations and pairings). Two out of three achieved reasonably fast running times, e.g. less than one second of computation for the client to retrieve 3MB from a 3GB-database, demonstrating the practicality of our proposal. 
\end{itemize}
We want to highlight another advantage of LMCs as the primitive for the committed PIR: as more advanced LMCs are developed, new features will be automatically added to the proposed scheme with no modifications to the underlying PIR schemes. For example, the LMC introduced in~\cite{CampanelliNitulescuRafolsZacharakisZapico_ePrint2022} allows \textit{updatability}, or the lattice-based LMC developed in~\cite{PeikertPepinSharp_TCC2021} provides \textit{post-quantum security}. 
 
The remainder of the paper is organized as follows. We first define formally the PIR and committed PIR schemes as well as their performance metrics in Section~\ref{sec:cpir}. We then introduce a generic construction for a committed PIR in Section~\ref{sec:generic}. In Section~\ref{sec:case_studies}, we demonstrate the proposed method with three case studies in which a linear map commitment is applied to three well-known PIR schemes.
Section~\ref{sec:exp} is devoted to implementations and evaluations.

\vspace{-10pt}

\section{Committed Private Information Retrieval}
\label{sec:cpir}

\subsection{Basic Notations}

We use $\bbF$ to denote a general finite field and $\fp$ to denote a finite field of $p$ elements, where $p$ is a prime power. Within the scope of this work, we usually assume that $p$ is a prime and hence, $\fp \equiv Z_p$, the integers mod $p$. For implementation, we use a specific prime $p$ of size about 256 bits (following the parameter of the BLS12-381 elliptic curve). We use $[n]$ to denote the set $\{1,2,\ldots,n\}$. The data is represented by $\bx = \xon$, where $x_j \in \ff^m$, $j \in [m]$, and $m\geq 1$ represents the size of each data item $x_j$ (in field elements).
We call $n$ the number of items or the size of the data. 
The data can also be regarded as an $m \times n$ matrix $\bx \in \fmn$ and each data item $x_j$ corresponds to the $j$-th column of the matrix.

Throughout this work we denote by $\lambda\in \bbN$ the security parameter, e.g. $\lambda = 128$, and $\negl(\lambda)$ the set of \textit{negligible functions} in $\lambda$.
A positive-valued function $\vep(\lambda)$ belongs to $\negl(\lambda)$ if for every $c>0$, there exists a $\lambda_0\in \bbN$ such that $\vep(\lambda)<1/\lambda^c$ for all $\lambda > \lambda_0$. We use $\poly(\lambda)$ for the set of polynomials in $\lambda$.

Before introducing the notation of a committed PIR scheme, we discuss the basic PIR and its performance metrics below. 

\vspace{-10pt}

\subsection{Private Information Retrieval}

A (replicated) PIR scheme has $k$ servers, each of which stores the data $\bx = (x_1,x_2,\ldots,x_n)$, and one client, who is interested in retrieving $x_i$ for some $i\in [n]$.

\begin{definition}[PIR]
\label{def:pir}
A $k$-server $n$-dimensional PIR scheme $\Pi_0$ over a field $\bbF$ consists of three algorithms $(\sf{QueriesGen},\sf{AnswerGen}, \sf{Extract})$ defined as follows.
\begin{itemize}
    \item $\big(\{q_j\}_{j\in [k]}, \aux \big) \leftarrow {\sf{QueriesGen}}(n, k, i)$: run by the client, this randomized algorithm takes as input $n>1$, $k\geq 1$, an index $i \in [n]$, and outputs $k$ queries to be sent to $k$ servers and an auxiliary information $\aux$.
    \item $a_j \leftarrow {\sf{AnswerGen}}(\bx, q_j)$: run by a server, this deterministic algorithm takes as input the data $\bx \in \bbF^n$, the query $q_j$, and outputs an answer $a_j$ to be sent to the client.
    \item $\{x_i\} \leftarrow {\sf{Extract}}\big(n, i, \{a_j\}_{j \in [k]}, \aux \big) $: run by the client, this deterministic algorithm takes as input $n$, $i$, the auxiliary information $\aux$, the answers from all $k$ servers, and outputs $x_i$.
\end{itemize}
\end{definition} 

A PIR scheme is called \textit{linear} if each answer $a_j$ is a linear combination of $\bx$.
We define below the correctness and privacy of a PIR scheme. 

\begin{definition}[Correctness of PIR] 
\label{def:correctness1}
The \kn PIR scheme defined in Definition~\ref{def:pir} is correct if for any $i\in[n]$, $\bx \in \bbF^n$, $\big(\{q_j\}_{j\in [k]}, {\sf{aux}}\big) \leftarrow \quegen(n, i)$, and $a_j \leftarrow \ansgen(\bx, q_j)$, $j\in [k]$, it holds that 
\[
\ext\big(n, i, \{a_j\}_{j \in [k]}, \aux \big) = x_i.
\]
\end{definition}

\begin{definition}[Privacy of PIR] The \kn PIR scheme defined in Definition~\ref{def:pir} is (unconditionally) $t$-private if no collusion of up to $t$ servers can learn any information about $i$, or more formally, for any $i, i'\in [n]$, and any subset $T\subsetneq [k]$ of size $|T|\leq t$, the distributions of $\quegen_T(n, k, i)$ and $\quegen_T(n, k, i')$ are identical, where $\quegen_T(n, k, i)$ denotes the concatenation of the $|T|$ output queries $\{q_j\}_{j \in T}$ generated by $\quegen(n, k, i)$. 
\end{definition}

\subsection{Communication and Computation Costs of PIR}

The efficiency of a PIR scheme can be measured based on its \textit{communication} and \textit{computation} costs. We first discuss the communication cost, which can be formally defined as follows.

\begin{definition}[Communication Cost of PIR]
\label{def:comm}
The communication cost of a PIR scheme $\Pi_0$ over a field $\ff$ given in Definition~\ref{def:pir} is defined as
\[
{\sf{comm}}(\Pi_0) = {\sf{up}}(\Pi_0) + {\sf{down}}(\Pi_0) \define \max_i\sum_{j \in [k]}|q_j| + 
\max_i\sum_{j \in [k]}|a_j|,
\]
where $|q_j|$ and $|a_j|$ denote the sizes (in field elements) of $q_j$ and $a_j$. The first term is the upload cost whereas the second is the download cost.
\end{definition}

For instance, in the aforementioned CKGS scheme~\cite{Chor-Kushilevitz-Goldreich-Sudan_JACM1998}, to represent a random subset of $\{1,2,\ldots,n\}$, the client must use a vector of $n$ bits, which means that the upload cost is $kn$ bits.
Straightforward generalizations of this scheme to $k>2$ servers (see, e.g.~\cite{DemmlerDaniel2014RPMP}) require an upload cost of $kn$ $\bbF$-elements, which is already significant for large $n$.
The main goal of the majority of early works on PIR was to optimize the communication cost.
The lowest known communication cost, namely, $O(kn^{1/d})$, for any $d \geq 1$, was achieved in the work of Woodruff and Yekhanin~\cite{WoodruffYekhanin_SJC2007}.
Their idea is to transform the PIR problem into the secret sharing problem while representing an index $i \in \{1,2,\ldots,n\}$ by a vector of length $O(n^{1/d})$ of Hamming weight $d$. We refer to this as the WY scheme.

\textbf{Download Rate.}
 Another approach to reduce the communication cost is to optimize the download cost, assuming that the data items are of large size and hence the upload cost will be overshadowed by the download cost (see, e.g. Sun and Jafar~\cite{SunJafar_TIT2018}).
 More precisely, one can aim for maximizing the \textit{download rate}, defined as $\max_{i\in [n]}\frac{|x_i|}{\sum_{j \in [k]}|a_j|}$, which is the ratio of the size of the desirable data to the total amount of data downloaded by the client. 
 Note that in the CKGS scheme, as the client downloads $k$ field elements from $k$ servers to recover one element, the download rate is $1/k$, which is quite small. 
 PIR schemes such as BE~\cite{BitarRouayheb_ITW2018} can achieve an asymptotically optimal rate of $(k-1)/k$.

\textbf{Computation Cost}. The computation cost of a PIR scheme typically consists of the computation time required by the client in generating the request and in recovering the desired data $x_i$, and the computation time required by the servers in producing the answers (taking the average or maximum among all servers). In general, as the client often has low computational capacity, its computation load, ideally, should be much less than that of the servers.

\vspace{-10pt}
\subsection{Committed Private Information Retrieval}

Apart from the large amount of research aiming for optimizing the upload or the download costs of a PIR scheme, there have also been a number of proposals in the literature that seek to extend the basic setting of the PIR problem (see, e.g. ~\cite{Ulukus_etal_JSAC2022} for a survey). 
In the scope of this work, we are interested in the \textit{verifiablity}, \textit{accountability}, and \textit{Byzantine-robustness} of a PIR scheme.

A $k$-server PIR scheme is \textit{$v$-verifiable} if the client can verify the correctness of the retrieved $x_i$ even when $v \leq k$ servers are colluding and try to fool the client by sending manipulated data. A scheme is \textit{$a$-accountable} if the client can identify 
all servers that sent incorrect data when at most $a\leq k$ servers did so.
A scheme is \textit{$b$-Byzantine-robust} if the client can recover the correct desired item $x_i$ when at most $b < k$ servers sent incorrect data.
It is clear that Byzantine-robustness implies accountability, which in turn implies verifiability. The converse is not true. However, it seems that a $b$-Byzantine-robust scheme can be obtained from a $b$-accountable scheme by increasing the number of servers communicated to obtain extra data for recovery (discarding the data received from identified malicious servers).
Readers who are familiar with coding theory may notice that the concepts of verifiability, accountability, and Byzantine-robustness defined above correspond to the classical concepts of \textit{error detection}, \textit{error-location identification}, and \textit{error correction}, respectively, in the study of channel coding. 

Following the notations of~\cite{ZhangSafaviNaini_ACNS2014}, we consider three types of participants: a \textit{data owner}\footnote{\scriptsize {In PIR's original setting, the servers are (implicitly) identical to the data owner. With the ubiquity of  cloud computing and the various benefits they offer, outsourcing storage/computing tasks to hired servers has become the trend. Thus, it is more practical to explicitly separate the data owner and the storage servers.}}, $k$ \textit{servers} $S_1,\ldots,S_k$, and a \textit{client}. 
The data owner owns the data~$\bx$. 
Although treated as a single trusted entity in theory, the data owner may also consist of multiple decentralized entities, e.g. a blockchain, which is maintained by a large number of miners.
Although each individual miner should not be trusted, the whole miner group are collectively trusted to produce valid commitments to the data, i.e., the block headers or the Merkle roots of transactions inside the block headers. The servers, on the other hand, are considered untrusted.

We formally define the Committed Private Information Retrieval (Com-PIR) scheme 
in Definition~\ref{def:cpir}. 
Compared to the basic PIR (see Definition~\ref{def:pir}), we also include one more dimension, $m$, to explicitly include the size of each data item. 

\begin{definition}[Com-PIR]
\label{def:cpir}
A \kmn committed PIR scheme 
$\Pi$ over a field $\ff$ consists of six algorithms 
defined as follows.
\begin{itemize}
    \item $\pp \leftarrow {\sf{Setup}}(1^\lambda, k, m, n)$: run by the data owner or a trusted setup\footnote{\scriptsize {In practice, a trusted setup can be run by a group of many participants (the power-of-$\tau$ ceremony~\cite{nikolaenko2022powers}), and as long as one person discards their piece of data, the secret key used in the setup remains secret and unrecoverable)}}, this randomized algorithm takes as input $\lambda$, $k, m, n$, where $\lambda$ is the security parameter, $k$ is the number of servers, $m$ is the size of each data item, $n$ is the number of data items, and outputs a public parameter $\pp$ known to everyone.
    \item $\cbx \leftarrow {\sf{CommitmentGen}}(\pp, \bx)$: run by the data owner, this deterministic algorithm takes as input the public parameter and the data $\bx \in \fmn$ and outputs the commitment $\cbx$.
    \item $\big(\{q_j\}_{j\in [k]}, \aux \big) \leftarrow {\sf{QueriesGen}}(\pp, k, m, n, i)$: run by the client, this randomized algorithm takes as input the public parameter, $k$, $m$, $n$, $i \in [n]$, and outputs $k$ queries to be sent to $k$ servers and an auxiliary information $\aux$.
    \item $a_j \leftarrow {\sf{AnswerGen}}(\pp, \bx, q_j)$: run by a server, this deterministic algorithm takes as input the public parameter, the data $\bx$, the query $q_j$, and outputs an answer $a_j$ to be sent to the client.
    \item $w_j \leftarrow {\sf{WitnessGen}}(\pp, \bx, q_j)$: run by a server, this deterministic algorithm takes as input the public parameter, the data $\bx$, and the query $q_j$, and outputs a witness $w_j$ to be sent to the client.
    \item $\{x_i,\perp\} \leftarrow {\sf{Extract}}\big(\pp, C, m, n, i, \{a_j,w_j\}_{j \in [k]}, \aux \big) $: run by the client, this deterministic algorithm takes as input the public parameter, a commitment $C$, $m$, $n$, $i\in [n]$, the answers and witnesses from all servers, the auxiliary information $\aux$, and outputs either $x_i$ (successful) or $\perp$ (unsuccessful).
    Note that $x_i$ denotes the $i$th column of the matrix $\bx$.
\end{itemize}
\end{definition}

A Com-PIR works as follows. First, the data owner or a trusted setup generates the public parameter $\pp$, which is available to everyone. Next, the data owner generates the commitment $\cbx$, which is made publicly available to everyone, e.g. by being embedded into a block header in a blockchain.
The client, who wants to retrieve $x_i$ privately, generates and sends queries to all servers. The servers generate and send the answers and the witnesses of the answers back to the client.
Finally, the client recovers $x_i$ and also performs the verification of the result using the commitment and the witnesses. 
The correctness, privacy, and verifiability of a Com-PIR scheme are formally defined below. 

\begin{definition}[Correctness of Com-PIR] 
\label{def:correctness2}
The \kmn Com-PIR scheme defined in Definition~\ref{def:cpir} is correct if the client can recover $x_i$ when all servers are honest, or more formally, for any $i\in[n]$, $\bx \in \fmn$, and $\pp \leftarrow \setup(1^\lambda, k, m, n)$, and $\big(\{q_j\}_{j\in [k]}, {\sf{aux}}\big) \leftarrow \quegen(\pp, k, m, n, i)$, and $a_j \leftarrow \ansgen(\pp, \bx, q_j)$, $w_j \leftarrow \witgen(\pp, \bx, q_j)$, $j\in [k]$, it holds that \vspace{-5pt} 
\[
\ext\big(\cbx, m, n, i, \{a_j,w_j\}_{j \in [k]}, \aux \big) = x_i.
\]
\end{definition}

\begin{definition}[Privacy of Com-PIR] The \kmn Com-PIR scheme defined in Definition~\ref{def:cpir} is (unconditionally) $t$-private if no collusion of up to $t$ servers can learn any information about $i$, or more formally, for any $i, i'\in [n]$, and any subset $T\subsetneq [k]$ of size $|T|\leq t$, the distributions of $\quegen_T(\pp, k, m, n, i)$ and $\quegen_T(\pp, k, m, n, i')$ are identical, where $\quegen_T(\pp, k, m, n, i)$ denotes the concatenation of the $|T|$ queries $\{q_j\}_{j \in T}$ output by $\quegen(\pp, k, m, n, i)$. 
\end{definition}

The \textit{verifiability} property of a Com-PIR is defined through the notion of a security experiment, in which an adversary $\A$ controls a group of Byzantine servers $\{S_j\}_{j \in B}$, $B\subseteq [k]$, knows the data $\bx$, the index $i$ (which means the privacy can be lost), and crafts the answers $\{\hat{a}_j\}_{j \in B}$ after receiving the queries $\{q_j\}_{j\in B}$. The goal of the adversary is to make the client accept an output $\xhi \notin \{x_i,\perp\}$.

\begin{definition}[Security Experiment for Verifiability]
The Com-PIR sche-me $\Pi$ defined in Definition~\ref{def:cpir} is $v$-verifiable if for any probabilistic polynomial time (PPT) adversary $\A$, there exists a negligible function $\vep(\lambda) \in \negl(\lambda)$ such that for any $i \in [n]$, any $\bx\in \fqmn$, and any subset $B\subseteq [k]$, $|B|\leq v$, it holds that 
\[
\pr\big[\EXP = 1 \big] \leq \vep(\lambda),
\]
where the security experiment $\EXP$ between an adversary and a challenger is described as follows.
\begin{itemize}
    \item The challenger picks $(\sk,\pp) \leftarrow {\sf{Setup}}(1^\lambda, k, m, n)$ and gives $\pp$ to $\A$.
    \item The adversary picks an $\bx \in \fqmn$, an $i \in [n]$, and a set $B\subseteq [k]$, $|B|\leq v$, and gives $\bx$, $i$, and $B$ to the challenger.
    \item The challenger generates $\cbx \leftarrow {\sf{CommitmentGen}}(\pp, \bx)$ and 
    $\big(\{q_j\}_{j\in [k]}, \aux \big)$ $\leftarrow {\sf{QueriesGen}}(\pp, k, m, n, i)$ and gives $\{q_j\}_{j\in B}$ to $\A$.
    \item The adversary crafts and gives $|B|$ answers and witnesses to the challenger \[\{\haj,\hwj\}_{j \in B}\leftarrow\A(\pp, k, \mathbf{x},i,B,\{q_j\}_{j\in B}).\]
    \item The challenger computes $\{a_j\}_{j \in [k] \setminus B} \leftarrow {\sf{AnswerGen}}(\pp, \bx, \{q_j\}_{j \in [k] \setminus B})$ and $\{w_j\}_{j \in [k] \setminus B} \leftarrow {\sf{WitnessGen}}(\pp, \bx, \{q_j\}_{j \in [k] \setminus B})$.
    \item The challenger runs the extraction algorithm 
    \[
    \xhi \leftarrow {\sf{Extract}}\big(\pp, \cbx, m, n, i, \{\haj,\hwj\}_{j \in B}, \{a_j,w_j\}_{j \in [k] \setminus B}, \aux \big).
    \]
    \item If $\xhi \notin\{x_i,\perp\}$ then set $\EXP=1$, and $0$, for otherwise.
\end{itemize}
\end{definition}

Note that to allow accountability and Byzantine-robustness for Com-PIR, one can include a set $B\subseteq [k]$ in the output of the algorithm $\extract(\cdot)$ to list identified Byzantine servers and then define corresponding security experiments. We omit the details and focus on verifiability only.

\vspace{-10pt}

\section{A Generic Construction of $k$-Verifiable Committed Private Information Retrieval Schemes}
\label{sec:generic}

We propose a generic construction for $k$-verifiable committed PIR schemes based on linear map commitment schemes and linear PIR schemes. The key idea is for the client to privately retrieve both $x_i$ and it hash $h_i$ using the same PIR scheme, where the correctness of the hash can be guaranteed by the linear map commitment. The client then verifies if the hash matches the data in the verification step. We first discuss the linear map commitment. 

\vspace{-10pt}
\subsection{Linear Map Commitments}
\label{subsec:LMC}

An $n$-dimensional linear map commitment allows a prover to first commit to a vector $\bx = (x_1,x_2,\ldots,x_n)$ and then prove to a verifier that a linear combination of $x_i$'s is correct, i.e. consistent with the commitment.
We formally define the linear map commitment schemes below, following~\cite{LaiMalavolta_CRYPTO2019}, \cite{Lai_Thesis_2022}.

\begin{definition}[Linear Map Commitments]
\label{def:LMC}
An $n$-dimensional linear map commitment (LMC) scheme $\Lambda$ over a field $\ff$ consists of four algorithms defined as follows.
\begin{itemize}
    \item $\pp \leftarrow {\sf{Setup}}(1^\lambda, n; \omega)$: this randomized algorithm takes as input $\lambda$, $n$, and $\omega$, where $\lambda$ is the security parameter, $n$ is the number of data items, $\omega$ is a random tape, and outputs a public parameter $\pp$ known to all parties. 
    \item $C \leftarrow {\sf{CommitmentGen}}(\pp, \bx)$: run by the prover, this deterministic algorithm takes as input the public parameter $\pp$ and the data $\bx = (x_1,x_2,\ldots,x_n)$ $\in \fn$ and outputs the commitment $C = \cbx$.
    \item $w_j \leftarrow {\sf{WitnessGen}}(\pp, \bx, \bc, y)$: run by a prover, this deterministic algorithm takes as input the public parameter $\pp$, the data $\bx = (x_1,x_2,\ldots,x_n)$ $\in \fn$, the vector of coefficients $\bc = (c_1,c_2,\ldots,c_n) \in \fn$, a value $y \in \ff$, and outputs a witness $w$ that proves that $y = \bc\cdot\bx=\sum_{j\in [n]} c_jx_j$.
    \item $\{0, 1\} \leftarrow {\sf{Verify}}\big(\pp, C, \bc, y, w \big)$: run by the verifier, this deterministic algorithm takes as input the public parameter $\pp$, a commitment $C$, a coefficient vector $\bc$, an element $y$, a witness $w$, and outputs either 1 or 0 to accept or reject that $y=\bc\cdot\bx$, respectively.
\end{itemize}
\end{definition}

There have been a few different constructions of LMC and variants/extensions recently proposed in the literature~\cite{LibertRamannaYung_LIPIcs16}, \cite{LaiMalavolta_CRYPTO2019}, \cite{Lai_Thesis_2022}, \cite{PeikertPepinSharp_TCC2021}, \cite{CampanelliNitulescuRafolsZacharakisZapico_ePrint2022}. The LMC in~\cite{LibertRamannaYung_LIPIcs16} is based on a ring and may not work immediately with a linear PIR scheme, which is often based on a finite field.
We use in this work the version of LMC introduced in the work of Lai and Malavolta~\cite{LaiMalavolta_CRYPTO2019}, \cite{Lai_Thesis_2022}, which is the most straightforward to implement and sufficient for our purpose. We refer to it as the Lai-Malavolta (LM) linear map commitment.
This LMC is based on an observation that the inner product of $\bc$ and $\bx$ is equal to the coefficient of $z^{n+1}$ in the product of the polynomials $f_{\bc}(z) \define \sum_{j\in [n]} c_j z^{n+1-j}$ and $f_{\bx}(z) \define \sum_{j \in [n]} x_j z^j$. 

\vspace{-10pt}
\begin{algorithm*}[htb!]
\begin{multicols}{2} 
\setup$(1^\lambda, n; \omega)$
\vspace{1pt}
\hrule
\vspace{1pt}
$\bg\leftarrow \bggen(1^\lambda; \omega)$\\
where $\bg \define (p, \bbG_1, \bbG_2, \bbG_T, G_1, G_2, e)$\\
$\al \leftarrow \bbZ_p$\\
$\pp = \big(\bg, \{G_1^{\alj}\}_{j \in [n]}, \{G_2^{\alj}\}_{j \in [2n]\setminus \{n+1\}}\big)$\\ 
\textbf{return} \pp\\

\comgen$(\pp, \bx)$
\vspace{2pt}
\hrule
\vspace{2pt}
\textbf{return} $C \define \prod_{j \in [n]} \big(G_1^{\alj}\big)^{x_j}$\\

\witgen$(\pp, \bx, \bc)$
\vspace{1pt}
\hrule
\vspace{1pt}
$w \define \prod_{j\in [n]}\prod_{j'\in [n]\setminus \{j\}} \big(G_2^{\al^{n+1+j-j'}}\big)^{c_jx_{j'}}$\\
\textbf{return } $w$\\

\verify$(\pp, C, \bc, y, w)$
\vspace{1pt}
\hrule
\vspace{1pt}
$b_0 \define \big(y \in \bbZ_p\big)$\\
$b_1 \define 
\begin{pmatrix} e\bigg(C, \prod_{j \in [n]} \big(G_2^{\al^{n+1-j}}\big)^{c_j}\bigg)\\
=e\bigg(\big(G_1^{\al}\big)^y, G_2^{\al^n}\bigg)e(G_1, w)
\end{pmatrix}$\\
\textbf{return $b_0$ AND $b_1$}
\end{multicols}
\vspace{-4pt}
\caption{The Lai-Malavolta linear map commitment scheme~\cite{LaiMalavolta_CRYPTO2019,Lai_Thesis_2022}.}
\end{algorithm*}

\vspace{-10pt}

\textbf{Lai-Malavolta Linear Map Commitment} (Algorithm~1). $\setup$ takes as input the security parameter $\lambda$, the vector length $n$, and a (private) random tape $\omega$ and outputs the public parameter $\pp$. First, \bggen{ } generates a bilinear group \bg, which includes a prime $p$, three cyclic groups $\bbG_1$, $\bbG_2$, $\bbG_T$ of order $p$ (written multiplicatively), where $G_1$ and $G_2$ are generators of $\bbG_1$ and $\bbG_2$, respectively, and $e\colon \bbG_1 \times \bbG_2 \to \bbG_T$ is a pairing satisfying the following properties:
\begin{itemize}
    \item $e$ is efficiently computable,
    \item $e$ is non-degenerate: $e(G_1,G_2) \neq 1_{\bbG_T}$,
    \item $e$ is bilinear: $e(A^x, B^y)=e(A,B)^{xy}$, for every $A\in \bbG_1$, $B\in \bbG_2$, and $x, y \in \bbZ$.
\end{itemize}
Next, a random element $\al$ is sampled from $\zp$. The output $\pp$ consists of the bilinear group, $\{G_1^{\alj}\}_{j \in [n]}$, and $\{G_2^{\alj}\}_{j \in [2n]\setminus \{n+1\}}$.
The commitment of $\bx$ and the witness for a linear combination $y=\bc\cdot \bx$ are computed as illustrated in Algorithm~1 (note that $y$ is not used in $\witgen$ in this scheme). Finally, $\verify$ checks if $y$ is an element in $\zp$ and verify if the first pairing is equal to the product of the other two. It accepts that $y=\bc\cdot \bx$ if both checks pass.

\textbf{Computational complexity of Lai-Malavolta LMC.} The LMC scales linearly for the verifier and quadratically in $n$ for the server. More specifically, in our implementation, the prover performs $O(n)$ elliptic curve operations and $O(n^2)$ field operations (cheaper) per linear combination. The verifier performs $O(n)$ elliptic curve operations and three pairings per linear combinations. Note that elliptic curve pairing $e(G,H)$ is more expensive than exponentiation $G^x$, which is more expensive than product $GH$, which in turn is more expensive than operations on finite fields. 
The Lai-Malavolta LMC requires a trusted setup and a linear-size public parameter, but provides a constant-size commitment and witness. Others constructions of LMC bring in different trade-offs, e.g. no trusted setup but $\log$-size commitment/witness, and additional properties~\cite{LibertRamannaYung_LIPIcs16,PeikertPepinSharp_TCC2021,CampanelliNitulescuRafolsZacharakisZapico_ePrint2022}.

\subsection{A Generic Construction of Com-PIR}
\label{subsec:generic}

\begin{figure}[htb!]
    \centering
\includegraphics[scale=0.88]{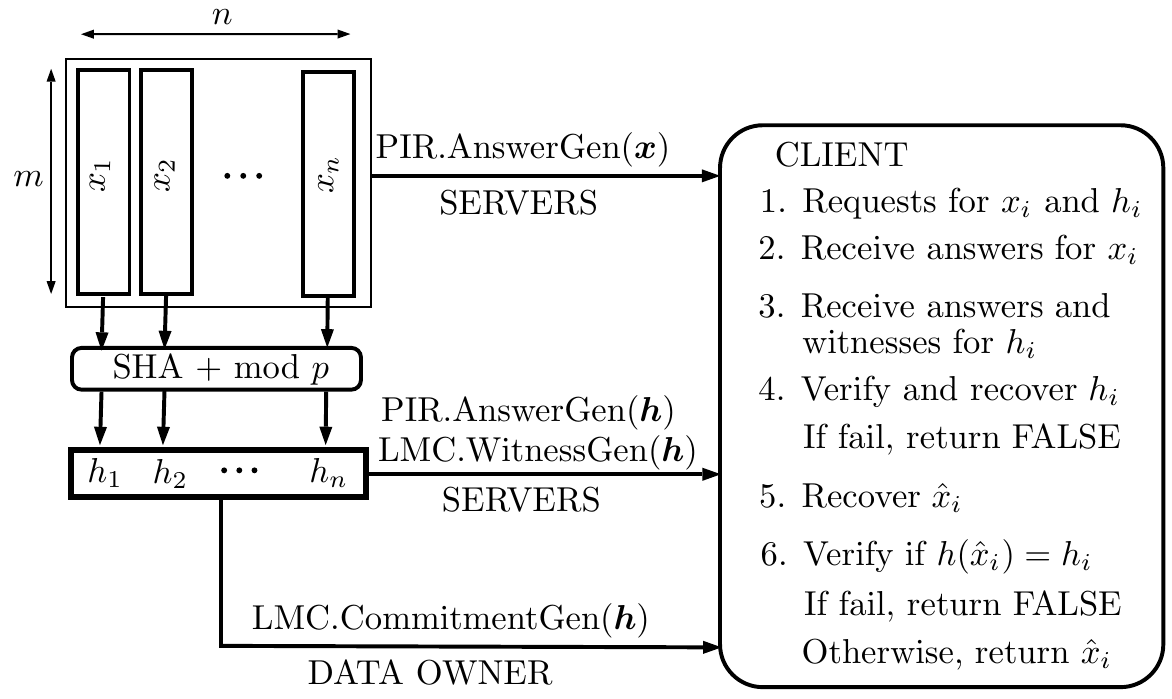}
    \vspace{-10pt}
    \caption{Illustration of a generic construction of Com-PIR using an LMC and a linear PIR. When using SHA3-256 and the BLS12-381, each $x_j$ is first hashed by SHA3-256 to generate a 256-bit digest $h^*(x_j)$, which in turn is taken modulo $p$, a 255-bit prime, to produce $h_j \in \zp$. This only reduces the security of SHA3-256 by at most one bit. The modulo operation is necessary to turn the hashes into valid input to the LMC. The LMC computation is carried out over the $n$ hashes only, making it independent of the data item size $m$.}
    \label{fig:cPIR}
\end{figure}

We now introduce a generic construction that combines an $n$-dimensional LMC and a linear \kmn PIR to produce a $k$-verifiable Com-PIR (see Fig.~\ref{fig:cPIR} for an illustration). The construction first applies a cryptographic hash function $h^*(\cdot)$ followed by a modulo operation to each column of the database $\bx$ to achieve $h_j = h^*(x_j) \pmod p$, where $x_j$ denotes the $j$th column of $\bx \in \fmn = \zp^{m\times n}$. It then applies an LMC to the vector $\bh = (h_1,h_2,\ldots,h_n)\in \ff^n$. The client performs PIR requests for \textit{both} $x_i$ and $h_i$. As the correctness of the received $h_i$ is guaranteed by the LMC, the verification $h_i \overset{?}{=} h(\hat{x}_i)$ is reliable.

\textbf{A Generic Construction of Com-PIR.} Let $\Pi$ be the target Com-PIR scheme, which will be constructed based on an LMC and a linear PIR scheme. We use the `.' sign to refer to the algorithm of each scheme, e.g. PIR.$\quegen()$.
\begin{itemize}
    \item $\pp \leftarrow \Pi.\setup(1^\lambda, k, m, n)$: The algorithm invokes $\lmc.\setup(1^\lambda, n)$, the setup algorithm of the LMC.
    \item $C \leftarrow \Pi.\comgen(\pp, \bx)$: The algorithm first computes $h_j = h^*(x_j) \pmod p, j \in [n]$, where $h^*(\cdot)$ is a cryptographic hash function, e.g. SHA3-256, and $p$ is the order of the cyclic groups as part of $\pp$. It then computes $C = C(\bh) \leftarrow \lmc.\comgen(\pp,\bh)$, where $\bh = (h_1,h_2,\ldots,h_n)$.
    \item $\big(\{q_j\}_{j\in [k]}, \aux \big) \leftarrow \Pi.\quegen(\pp, k, m, n, i)$: The algorithm invokes the corresponding PIR algorithm, namely, $\pir.\quegen(\pp,k,m,n,i)$.
    \item $a_j \leftarrow \Pi.\ansgen(\pp, \bx, q_j)$: The algorithm invokes the corresponding PIR algorithm on both $\bx$ and $\bh$, i.e. $a_j(\bx) \leftarrow \pir.\ansgen(\pp,\bx,q_j)$ and $a_j(\bh) \leftarrow \pir.\ansgen(\pp,\bh,q_j)$, and outputs $a_j \define (a_j(\bx),a_j(\bh))$. Note that each server can compute $\bh$ from $\bx$ on its own just once. 
    \item $w_j \leftarrow \Pi.\witgen(\pp, \bx, q_j)$: The algorithm first converts $q_j$ into a coefficient vector $\bc(q_j) \in \zp^n$ such that $a_j(\bh) = \bc(q_j)\cdot \bh$. Then, it invokes $w_j \leftarrow \lmc.\witgen(\pp,\bh,\bc,y)$, where $y \define \bc\cdot \bh$.
    \item $\{x_i,\perp\} \leftarrow \Pi.\extract\big(\pp, C, m, n, i, \{a_j,w_j\}_{j \in [k]}, \aux \big) $: The algorithm first parses each answer $a_j$ as $\big(a_j(\bx), a_j(\bh)\big)$. Next, it converts $q_j$ into a coefficient vector $\bc(q_j) \in \zp^n$ such that $a_j(\bh) = \bc(q_j)\cdot \bh$. Then, it verifies $a_j(\bh)$ by running $\lmc.\verify(\pp,C,\bc(q_j),a_j(\bh), w_j), j \in [k]$. 
    If the verification fails for $j \in [k]$, it returns $\perp$. Otherwise, it calls $\pir.\extract\big(n,i,\{a_j(\bh)\}_{j\in [k]}, \aux\big)$ and $\pir.\extract\big(n,i,\{a_j(\bx)\}_{j \in [k]}, \aux\big)$ to obtain $\hhi$ and $\xhi$. It then performs the final hash verification $h(\xhi) \overset{?}{=} \hhi$ and returns $\xhi$ if passes and $\perp$ if fails.
\end{itemize}

Note that in the generic construction above, the PIR scheme is applied to $\bx\in \zp^{m\times n}$ instead of $\zp^n$ as in Definition~\ref{def:pir}. This can be done in a straightforward manner in which the PIR scheme on $\zp^n$ is applied repeatedly $m$ times to the $m$ rows of $\bx\in \zp^{m\times n}$ using the same set of queries. 
The communication and computation costs of a Com-PIR scheme based on the generic construction can be calculated easily based on the costs of the underlying PIR and LMC. 
Note that the LMC doesn't depend on $m$. Although the final hash check $h(\xhi) \overset{?}{=} \hhi$ depends on $m$ because $|\xhi|=m$, $h(\cdot)$ is very efficient and its cost is negligible.

\begin{lemma}[Correctness/Privacy]
The Com-PIR constructed by the generic construction is correct and $t$-private if the underlying LMC scheme is correct and the underlying PIR scheme is both correct and $t$-private.
\end{lemma}

\textit{Proof. }
The correctness of the constructed Com-PIR scheme can be proved in a straightforward manner, implied directly from the correctness of the underlying LMC and PIR schemes. The privacy of the Com-PIR follows from the privacy of the underlying PIR scheme because the queries sent from the client are identical to those in the original PIR scheme.

Next, we prove (see Appendix~A) that the generic construction generates a \kmn Com-PIR that is $k$-verifiable, assuming that the Lai-Malavolta LMC is used in conjunction with an arbitrary linear PIR scheme.

\begin{lemma}[Verifiability]
\label{lem:verifiability}
Let $k, m, n \in \poly(\lambda)$ and $1/p \in \negl(\lambda)$. Then the \kmn Com-PIR using the Lai-Malavolta LMC is $k$-verifiable in the generic bilinear group model.
\end{lemma}

\vspace{-10pt}

\section{Three Case Studies}
\label{sec:case_studies}
We discuss in detail how the generic construction proposed in Section~\ref{subsec:generic} performs for the Lai-Malavolta LMC and the three representative linear PIR schemes with respect to the communication and computation costs.

\textbf{Chor-Kushilevitz-Goldreich-Sudan (CKGS) Scheme}~(\cite{Chor-Kushilevitz-Goldreich-Sudan_JACM1998}). This is a linear 2-server $n$-dimensional PIR scheme working over an arbitrary finite field~$\bbF$.
We also use 2-CKGS to refer to this scheme, while using $k$-CKGS to refer to its straighforward generalization to the $k$-server setting (see \cite[Section~3.2.1]{DemmlerDaniel2014RPMP}).
\begin{itemize}
    \item $\big(\{q_1,q_2\}, \aux \big) \leftarrow {\sf{QueriesGen}}(n, 2, i)$: the algorithm first picks a random subset $J \subseteq [n]$ and let $q_1\in \bbF^n$ be the characteristic vector for $J$, i.e., $q_1$ has a `1' at the the $j$th component if $j \in J$, and $0$ otherwise. Next, $q_2$ is obtained from $q_1$ by flipping its $i$th component ($0\to 1$ or $1 \to 0$). Then either $e_i = q_1-q_2$ or $e_i = q_2-q_1$, where $e_i\in \ff^n$ is the unit vector with a `1' at the $i$th component. Set $\aux=1$ or $\aux = 2$, respectively.
    \item $a_j \leftarrow {\sf{AnswerGen}}(\bx, q_j)$: The algorithm returns $a_j = q_j \cdot \bx$.
    \item $\{x_i\} \leftarrow {\sf{Extract}}\big(n, i, \{a_j\}_{j \in [k]}, \aux \big) $: The algorithm returns $a_1-a_2$ if $\aux=1$ or $a_2-a_1$ if $\aux=2$. 
\end{itemize}
The CKGS scheme and the Lai-Malavolta LMC scheme work together in a straightforward manner. 

\begin{example}
\label{ex:LMC_Chor}
We consider in Fig.~\ref{fig:toy1} a toy example of a 2-server $m\times 5$-dimensional Com-PIR based on an LMC and CKGS PIR scheme~\cite{Chor-Kushilevitz-Goldreich-Sudan_JACM1998} (see Fig.~\ref{fig:toy1} for an illustration).
The client, who wants $x_1$, picks a random subset $J = \{1,3,4\}\subseteq [5]$ and creates the corresponding queries $q_1 = (1,0,1,1,0)$ and $q_2 = (0,0,1,1,0)$. Server 1, if acting honestly, sends back the answers $a_1(\bh) = h_1+h_3+h_4$, $a_1(\bx)=x_1+x_3+x_4$, and the witness $w_1(a_1(\bh))$, which allows the client to verify the correctness of $a_1(\bh)$. Server 2, if acting honestly, sends back the answers $a_2(\bh) = h_3+h_4$, $a_2(\bx)=x_3+x_4$, and the witness $w_2(a_2(\bh))$, which allows the client to verify the correctness of $a_2(\bh)$. The client, knowing the LMC commitment $C(\bh)$, can verify the correctness of both $a_1(\bh)$ and $a_2(\bh)$ and then extract the (verifiably correct) $\hat{h}_1 = a_1(\bh)-a_2(\bh)$. It can also extract $\hat{x}_1 = a_1(\bx)-a_2(\bx)$ and verify the correctness of the result by performing a hash verification $h(\hat{x}_1) \overset{?}{=} \hat{h}_1$.
\end{example}

\vspace{-5pt}
\textbf{Woodruff-Yekhanin (WY) Scheme}~(\cite{WoodruffYekhanin_SJC2007}). This is a linear $k$-server $n$-dimensional PIR scheme working over an arbitrary finite field $\bbF$. See Appendix~B for the detail.
Note that each server generates $\ell\in O(n^{1/d})$ linear combinations of $\bx$. Although in Algorithm~1, we let the LMC verify just one linear combination of $\bx$ for simplicity, in its original form, Lai-Malavolta LMC can produce a single witness for and verify multiple linear combinations. 

\textbf{Bitar-El Rouayheb (BE) Scheme}~(\cite{BitarRouayheb_ITW2018}). We present a simplified version of this scheme (dropping unnecessary properties like universality) in Appendix~B (see, also Goldberg~\cite{Goldberg_SP2007}).
Note that the LMC is applied on $(k-t)n$ hashes instead of $n$ like in other schemes.

\vspace{-10pt}
\begin{table}[hbt!]
\begin{center}
{\scriptsize
\begin{tabular}{|p{1.5cm}|p{1.8cm}|p{1.4cm}|p{3.1cm}|p{3.9cm}|}
\hline
& \textbf{Upload Cost} (\#$\zp$-elts) & \textbf{Download Rate}  & \textbf{Server} (\#operations) & \textbf{Client}  (\#operations)\\
\hline 
\hline \hline
\bfseries{Com-PIR} & $2n$ bits & $1/2$ & $mn+$ & $m+$\\
\cline{2-5}
 \bfseries{(2-CKGS)} & & & \vspace{0.1pt}$n^2+$, $n^2\times$, $n\boxplus$, $n\boxtimes$ & \vspace{0.1pt} $2n\boxplus$, $2n\boxtimes$, $6\boxdot$\\
\hline \hline
\bfseries{Com-PIR} & $kn$ & $1/k$ &  $mn+$, $mn\times$ &  $km+$\\
\cline{2-5}
 \bfseries{($k$-CKGS)} & & & \vspace{0.1pt} $n^2+$, $n^2\times$, $n\boxplus$, $n\boxtimes$ & \vspace{0.1pt} $kn\boxplus$, $kn\boxtimes$, $3k\boxdot$\\
\hline \hline

\bfseries{Com-PIR} & \vspace{0.1pt} $k\ell$ & \vspace{0.1pt} $1/k$ & \vspace{0.1pt}$\ell mn+$, $\ell mnd\times$ & \vspace{0.1pt}$mt(k \ell + d^3t^2)+$, $mt(k \ell + d^3t^2)\times$\\
\cline{2-5}
 \bfseries{(WY)} & & & \vspace{0.1pt} $\ell n^2+$, $\ell n^2\times$, $\ell n\boxplus$, $\ell n\boxtimes$ & \vspace{0.1pt}$k\ell n\boxplus$, $k\ell n\boxtimes$, $3k \boxdot$\\
\hline \hline

\bfseries{Com-PIR} & \vspace{0.1pt} $k(k-t)n$ &  \vspace{0.1pt} $(k-t)/k$ & \vspace{0.1pt}$(k-t)mn+$, $(k-t)mn\times$ & \vspace{0.1pt}
$k((k-t)(kn+m) + km)+$, $k^2((k-t)n + m)\times$\\
\cline{2-5}
 \bfseries{(BE)} & & & \vspace{0.1pt} $(k-t)^2n^2+$, $(k-t)^2n^2\times$, $(k-t)n\boxplus$, $(k-t)n\boxtimes$ & \vspace{0.1pt} $k(k-t)n\boxplus$, $k(k-t)n\boxtimes$, $3k\boxdot$\\
\hline
\end{tabular}}
\end{center}
\caption{Comparisons of different Com-PIR schemes with $k$-verifiability. For the computation costs at each server and client, we count the number of field additions `$+$' and multiplications `$\times$', elliptic curve additions `$\boxplus$' and multiplications `$\boxtimes$', and pairings `$\boxdot$' in big-O notation (with $\ell\in O(n^{1/d})$ for the WY-based scheme). The top sub-row counts the operations on \textit{data} while the bottom sub-row counts the operations related to \textit{verification}. The verification time of the proposed Com-PIR schemes (mostly) doesn't depend on the size of the retrieved data $m$ but on the size of the database $n$ and the number of servers $k$.}

\label{tab:comparison}
\end{table}
\vspace{-20pt}

\textbf{Comparison of the three Com-PIR schemes.} We compare these schemes based on their communication and computation complexities (Table~\ref{tab:comparison}).
\begin{itemize}
    \item \textbf{LM-CKGS}: this scheme has the \textit{lowest computation time} among the three for both servers and client. The reason is that each server only performs cheap field additions for the data part and generates LMC witness for a single linear combination of hashes. The client performs one LMC verification per server.
    \item \textbf{LM-WY}: although having the \textit{lowest upload cost}, the computation cost of this scheme is the highest among the three. 
    The reason is that its running time also depends on $\ell\in O(n^{1/d})$. See Appendix~C for more details.
    \item \textbf{LM-BE}: this scheme achieves the \textit{optimal download rate} and has computation cost lying in between the other two. Computation-wise, the smaller the difference $k-t$, the lower the running time of both servers and client. The reason is that the LMC has to run not on $n$ but on $(k-t)n$ hashes.
\end{itemize}

\textbf{Comparisons with related works.} 
The work of Zhang-Safavi Naini~\cite{ZhangSafaviNaini_ACNS2014} is the closest to ours and provides $k$-accountability. Their idea is to apply a verifiable computing scheme~\cite{PapamanthouShiTamassia_2013} on top of WY~\cite{WoodruffYekhanin_SJC2007}, followed by several optimization steps to improve the performance. Originally designed for a $1\times n$ database ($m = 1$), the verification time of their main scheme $\Gamma_1$ is in $O\big(kmn^{1/d}\big)$, which becomes very slow for large $m$.
Moreover, while our scheme has a constant witness size (from each server), their witness size is in $O\big(mn^{1/d}\big)$.
All other works in the literature, to our best knowledge, do not provide $k$-verifiability. For instance, Ke and Zhang~\cite{KeZhang_ISIT2022} constructed a 2-server PIR scheme that can (information theoretically) verify the correctness of the result given at most one malicious server. Zhang and Wang~\cite{ZhangWang_SP2022} introduced $k$-server PIR schemes that are privately and publicly $v$-verifiable for $v<k$. These schemes were also designed for $m = 1$. Zhao \textit{et al.}~\cite{ZhaoWangHuang_IS2021} proposed a construction of verifiable PIR scheme based on the Learning with Errors problem. The main issue in their construction is that the server can pass the client's verification if using the same \textit{incorrect} database in generating the answer for the query and the response to the challenge (see~\cite[Def.~7]{ZhaoWangHuang_IS2021}). 
PIR schemes with Byzantine-robustness were investigated in~\cite{DevetGoldbergHeninger_UsenixSS12,YangBennett_AICSA2002,BeimelStahl_SCN2002,Goldberg_SP2007}.

\vspace{-10pt}
\section{Experiments \& Evaluations}
\label{sec:exp}

\textbf{Experiment setup.}
We implemented three Com-PIR schemes in C using the libraries \texttt{GMP} 6.2.1, \texttt{OpenSSL} 2022, and \texttt{blst} v.0.3.10. 
We compiled the code with GCC 11.3.0 and  ran our experiments on Ubuntu 22.04.1 environment (Intel Core i5-1035G1 CPU @1.00GHz×8, 15GB System memory).  The code is available on GitHub at https://github.com/PIR-PIXR/CPIR.

\begin{figure} [hbt!]
    \centering
    \hspace{-12pt}
    \includegraphics[scale=0.4]{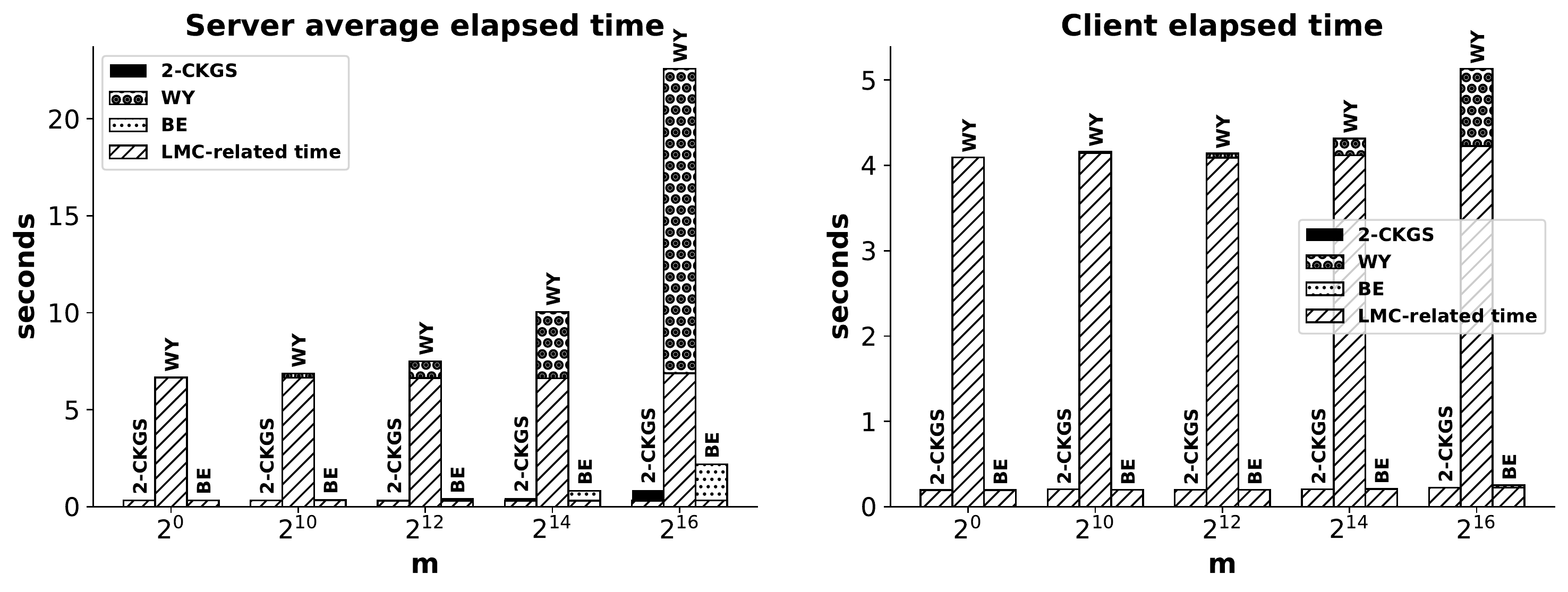}
    \vspace{-15pt}
    \caption{The average server and client computation times of LM-CKGS, LM-WY, and LM-BE for $k = 2$, $t = 1$, $n = 2^{10}$, and $m\in \{2^{0}, 2^{10}, 2^{12}, 2^{14}, 2^{16}\}$.}
    \label{fig: case1}
\end{figure}

\begin{figure} [hbt!]
    \centering
    \hspace{-11pt}
    \includegraphics[scale=0.4]{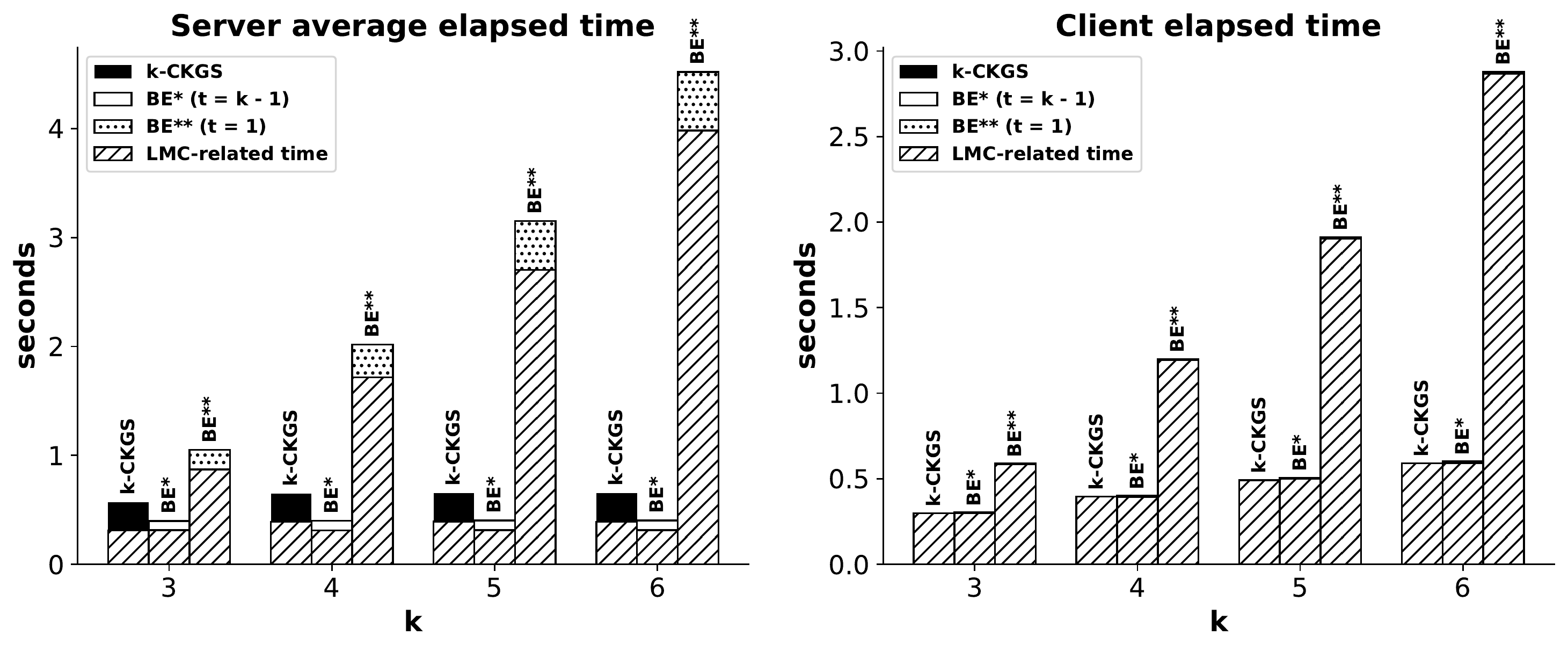}
    \vspace{-15pt}
    \caption{The average server and client computation times of LM-CKGS and LM-BE for $n = 2^{10}$, $m = 2^{12}$, $k\in \{3,4,5,6\}$, and $t = 1$ or $k - 1$.}
    \label{fig: case4}
\end{figure}

\textbf{Evaluations.}
The LMC component in the Com-PIR schemes incurs an extra communication/computation overhead on top of the original PIR schemes~\cite{Chor-Kushilevitz-Goldreich-Sudan_JACM1998,WoodruffYekhanin_SJC2007,BitarRouayheb_ITW2018}.
However, the LMC communication overhead is only $O(k)$ while the computation overhead doesn't grow with $m$, the size of each data item. Hence, as the size of each data item increases, the LMC overhead becomes smaller and smaller compared to the computation time of the PIR scheme (see Fig.~\ref{fig: case1}). The computation time of LM-WY is significantly higher than the other two schemes (see Fig.~\ref{fig: case1}, Fig.~\ref{fig: case4}), hence consistent with the theoretical analysis presented earlier. 
Fig.~\ref{fig: case4} also demonstrates a trade-off between the download rate and the computation time for LM-BE: larger $t$ leads to smaller download rate but cheaper computation.
More evaluations of these Com-PIR schemes are in Appendix~C. 

\vspace{-10pt}

\section{Conclusions}
\label{sec:conclusions}

We proposed a modular approach to combine a linear map commitment and a linear PIR scheme to achieve a $k$-verifiable PIR scheme, which guarantees that the client will never accept wrong data even in the extreme case when all servers are malicious. By applying the commitment scheme on hashes of data rather than on data themselves, the construction is reasonably practical, taking less than one second to privately retrieve 3MB of data from a database of size 3GB. A drawback of our approach is that the commitment scheme may incur a significant computation overhead on top of the PIR scheme if the database consists of a large number of small-sized items. 

\vspace{-10pt}
\section*{Acknowledgement}

We thank Russell W. F. Lai and Liangfeng Zhang for helpful discussions. 

\vspace{-10pt}
\bibliographystyle{IEEEtran}
\bibliography{CommittedPIR.bib}

\begin{thebibliography}{10}
\providecommand{\url}[1]{#1}
\csname url@samestyle\endcsname
\providecommand{\newblock}{\relax}
\providecommand{\bibinfo}[2]{#2}
\providecommand{\BIBentrySTDinterwordspacing}{\spaceskip=0pt\relax}
\providecommand{\BIBentryALTinterwordstretchfactor}{4}
\providecommand{\BIBentryALTinterwordspacing}{\spaceskip=\fontdimen2\font plus
\BIBentryALTinterwordstretchfactor\fontdimen3\font minus
  \fontdimen4\font\relax}
\providecommand{\BIBforeignlanguage}[2]{{%
\expandafter\ifx\csname l@#1\endcsname\relax
\typeout{** WARNING: IEEEtran.bst: No hyphenation pattern has been}%
\typeout{** loaded for the language `#1'. Using the pattern for}%
\typeout{** the default language instead.}%
\else
\language=\csname l@#1\endcsname
\fi
#2}}
\providecommand{\BIBdecl}{\relax}
\BIBdecl

\bibitem{UribeAlice2022CoOP}
A.~Uribe, ``Cyberattack on optus potentially exposes millions of customer
  accounts; australian telecoms company says mass breach could have exposed
  birth dates, phone numbers and other personal data,'' \emph{The Wall Street
  journal. Eastern edition}, 2022.

\bibitem{biddle2022public}
N.~Biddle, M.~Gray, and S.~McEachern, ``Public exposure and responses to data
  breaches in australia: October 2022,'' 2022.

\bibitem{Chor-Kushilevitz-Goldreich-Sudan_JACM1998}
B.~Chor, E.~Kushilevitz, O.~Goldreich, and M.~Sudan, ``Private information
  retrieval,'' \emph{Journal of the ACM (JACM)}, vol.~45, no.~6, pp. 965--981,
  1998.

\bibitem{KeZhang_ISIT2022}
P.~Ke and L.~F. Zhang, ``Two-server private information retrieval with result
  verification,'' in \emph{Proceedings of the IEEE International Symposium on
  Information Theory (ISIT)}, 2022, pp. 408--413.

\bibitem{ZhaoWangHuang_IS2021}
L.~Zhao, X.~Wang, and X.~Huang, ``Verifiable single-server private information
  retrieval from lwe with binary errors,'' \emph{Information Sciences}, vol.
  546, pp. 897--923, 2021.

\bibitem{ZhangWang_SP2022}
L.~F. Zhang and H.~Wang, ``Multi-server verifiable computation of low-degree
  polynomials,'' in \emph{Proceedings of the IEEE Symposium on Security and
  Privacy (S\&P)}, 2022, pp. 596--613.

\bibitem{DevetGoldbergHeninger_UsenixSS12}
C.~Devet, I.~Goldberg, and N.~Heninger, ``Optimally robust private information
  retrieval,'' in \emph{Proceedings of the 21st USENIX Security Symposium},
  2012, pp. 269--283.

\bibitem{YangBennett_AICSA2002}
E.~Y. Yang, J.~Xu, and K.~H. Bennett, ``Private information retrieval in the
  presence of malicious failures,'' in \emph{Proceedings of the 26th Annual
  International Computer Software and Applications}, 2002, pp. 805--810.

\bibitem{BeimelStahl_SCN2002}
A.~Beimel and Y.~Stahl, ``Robust information-theoretic private information
  retrieval,'' in \emph{Proceedings of the Third International Conference on
  Security in Communication Networks}, 2003, pp. 326--341.

\bibitem{BeimelStahl_JC2007}
------, ``Robust information-theoretic private information retrieval,''
  \emph{Journal of Cryptology}, vol.~20, pp. 295--321, 2007.

\bibitem{Goldberg_SP2007}
I.~Goldberg, ``Improving the robustness of private information retrieval,'' in
  \emph{2007 IEEE Symposium on Security and Privacy (SP'07)}.\hskip 1em plus
  0.5em minus 0.4em\relax IEEE, 2007, pp. 131--148.

\bibitem{ZhangSafaviNaini_ACNS2014}
L.~F. Zhang and R.~Safavi-Naini, ``Verifiable multi-server private information
  retrieval,'' in \emph{Proceedings of the International Conference on Applied
  Cryptography and Network Security (ANCS)}, 2014, pp. 62--79.

\bibitem{LaiMalavolta_CRYPTO2019}
R.~W. Lai and G.~Malavolta, ``Subvector commitments with application to
  succinct arguments,'' in \emph{Advances in Cryptology--CRYPTO 2019: 39th
  Annual International Cryptology Conference, Santa Barbara, CA, USA, August
  18--22, 2019, Proceedings, Part I 39}.\hskip 1em plus 0.5em minus 0.4em\relax
  Springer, 2019, pp. 530--560.

\bibitem{LibertRamannaYung_LIPIcs16}
B.~Libert, S.~C. Ramanna \emph{et~al.}, ``Functional commitment schemes: From
  polynomial commitments to pairing-based accumulators from simple
  assumptions,'' in \emph{43rd International Colloquium on Automata, Languages
  and Programming (ICALP 2016)}, 2016.

\bibitem{PeikertPepinSharp_TCC2021}
C.~Peikert, Z.~Pepin, and C.~Sharp, ``Vector and functional commitments from
  lattices,'' in \emph{Theory of Cryptography: 19th International Conference,
  TCC 2021, Raleigh, NC, USA, November 8--11, 2021, Proceedings, Part III
  19}.\hskip 1em plus 0.5em minus 0.4em\relax Springer, 2021, pp. 480--511.

\bibitem{CampanelliNitulescuRafolsZacharakisZapico_ePrint2022}
M.~Campanelli, A.~Nitulescu, C.~R{\`a}fols, A.~Zacharakis, and A.~Zapico,
  ``Linear-map vector commitments and their practical applications,'' in
  \emph{International Conference on the Theory and Application of Cryptology
  and Information Security}.\hskip 1em plus 0.5em minus 0.4em\relax Springer,
  2022, pp. 189--219.

\bibitem{WoodruffYekhanin_SJC2007}
D.~Woodruff and S.~Yekhanin, ``A geometric approach to information-theoretic
  private information retrieval,'' in \emph{20th Annual IEEE Conference on
  Computational Complexity (CCC'05)}.\hskip 1em plus 0.5em minus 0.4em\relax
  IEEE, 2005, pp. 275--284.

\bibitem{BitarRouayheb_ITW2018}
R.~Bitar and S.~El~Rouayheb, ``Staircase-pir: Universally robust private
  information retrieval,'' in \emph{2018 IEEE Information Theory Workshop
  (ITW)}.\hskip 1em plus 0.5em minus 0.4em\relax IEEE, 2018, pp. 1--5.

\bibitem{DemmlerDaniel2014RPMP}
D.~Demmler, A.~Herzberg, and T.~Schneider, ``Raid-pir: practical multi-server
  pir,'' in \emph{Proceedings of the 6th edition of the ACM Workshop on Cloud
  Computing Security}, 2014, pp. 45--56.

\bibitem{SunJafar_TIT2018}
H.~Sun and S.~A. Jafar, ``The capacity of robust private information retrieval
  with colluding databases,'' \emph{IEEE Transactions on Information Theory},
  vol.~64, no.~4, pp. 2361--2370, 2017.

\bibitem{Ulukus_etal_JSAC2022}
S.~Ulukus, S.~Avestimehr, M.~Gastpar, S.~A. Jafar, R.~Tandon, and C.~Tian,
  ``Private retrieval, computing, and learning: Recent progress and future
  challenges,'' \emph{IEEE Journal on Selected Areas in Communications},
  vol.~40, no.~3, pp. 729--748, 2022.

\bibitem{nikolaenko2022powers}
V.~Nikolaenko, S.~Ragsdale, J.~Bonneau, and D.~Boneh, ``Powers-of-tau to the
  people: Decentralizing setup ceremonies,'' \emph{Cryptology ePrint Archive},
  2022.

\bibitem{Lai_Thesis_2022}
R.~W. Lai, ``Succinct arguments: Constructions and applications,'' Ph.D.
  dissertation, Friedrich-Alexander-Universitaet Erlangen-Nuernberg (Germany),
  2022.

\bibitem{PapamanthouShiTamassia_2013}
C.~Papamanthou, E.~Shi, and R.~Tamassia, ``Signatures of correct computation,''
  in \emph{Theory of Cryptography Conference}.\hskip 1em plus 0.5em minus
  0.4em\relax Springer, 2013, pp. 222--242.

\end{thebibliography}

\vspace{-10pt}
\appendix
\section*{Appendix A: Verifiability Proof of Com-PIR}
\label{app:commit-to-data}

For simplicity, we consider one linear combination of $\bx$ in the definition below.

\begin{definition}[Function Binding for LMC]\cite{LaiMalavolta_CRYPTO2019}
\label{def:binding}
An LMC over $\ff$ is function binding if for any PPT adversary $\A$, any positive integer $n\in \poly(\lambda)$, there exists a negligible function $\varepsilon(\lambda)$ such that
\[
\pr
\left[
\begin{array}{c|c}
y \in \ff & \omega\mbox{ } \rand \mbox{ }\{0,1\}^{\lambda} \\
\mbox{ }\verify(C, \bc, y, w) = 1 & \mbox{ }\pp \leftarrow \setup(1^{\lambda}, n; \omega) \\
\not\exists \bx \in \ff^n \text{ s.t. } \sum_{j \in [n]} c_jx_j = y & (C, \bc, y, w) \leftarrow \A(\pp) \\
\end{array}
\right]
\leq \varepsilon(\lambda).
\]
\end{definition}

\begin{lemma}[\cite{LaiMalavolta_CRYPTO2019}]
\label{lem:binding}
Let $n \in \poly(\lambda)$ and $1/p \in \negl(\lambda)$. Then Lai-Malavolta LMC is function binding in the generic bilinear group model.
\end{lemma}

\textit{Proof. }[Proof of Lemma~\ref{lem:verifiability}]
According to the generic construction, the adversary wins the security experiment, i.e., $\EXP = 1$, if and only if the challenger extracts $\xhi \notin \{x_i,\perp\}$. This happens only when one of the following two independent events occur: either $\hhi = h_i$, i.e., the adversary finds a hash collision $h(\xhi)=h(x_i)=h_i$, or $\hhi \neq h_i$ but the adversary manages to fool $\lmc.\verify()$ with at least one wrong linear combination of $h_i$'s. Therefore,
\[
\begin{split}
&\pr[\EXP = 1] = \pr[\xhi \neq x_i \wedge h(\xhi)=h(x_i)]\\ &+ \pr[\text{At least one linear combination is wrong but still passes LMC.\verify()}]\\
&\leq \varepsilon_1(\lambda) + \varepsilon_2(\lambda) \in \negl(\lambda),\vspace{-5pt}
\end{split}
\]
as both events happen with probabilities negligible in $\lambda$ assuming that $1/p$ is negligible in $\lambda$ 
and that the security is considered under the generic bilinear group model (Lemma~\ref{lem:binding}).

\vspace{-10pt}
\section*{Appendix B: WY and BE PIR Schemes} 

\textbf{Woodruff-Yekhanin (WY) Scheme}~(\cite{WoodruffYekhanin_SJC2007}). 
This is a linear $k$-server $n$-dimensi-onal PIR scheme working over an arbitrary finite field $\bbF$. 
Let $1\leq t < k$ and $d = \lfloor (2k-1)/t\rfloor$. Let $\ell\in O\big(n^{1/d}\big)$ be the smallest integer satisfying $\binom{\ell}{d} \geq n$ and $E \colon [n] \to \fl$ a 1-to-1 mapping that maps an index $j\in [n]$ to a vector in $\fl$ of Hamming weight $d$. Each $\bx \in \fn$ is encoded by a multivariate polynomial $F_{\bx}(\bz)$, where $\bz = (z_1,\ldots,z_{\ell})$, defined as follows. \vspace{-5pt}
\[
F_{\bx}(\bz) \define \sum_{j \in [n]} x_j \prod_{u \in [\ell]\colon E(j)_u=1}z_u. \vspace{-5pt}
\]
Then, $\deg(F)=d$ and $x_i = F_{\bx}(E(i))$. Fix $k$ distinct elements $\{\beta_j\}_{j \in [k]}\subseteq \fp^*$.
\begin{itemize}
    \item $\big(\{q_j\}_{j\in [k]}, \aux \big) \leftarrow {\sf{QueriesGen}}(n, k, i)$: The algorithm picks $t$ random vectors $\{\vs\}_{s \in [t]}\subseteq \fl$ and outputs $q_j \define E(i)+\sum_{s\in [t]}\beta_j^sv_s$ and $\aux = \{\vs\}_{s \in [t]}$.
    \item $a_j \leftarrow {\sf{AnswerGen}}(\bx, q_j)$: The algorithm computes $a_j = (a_{j,0},a_{j,1},\ldots,a_{j,\ell})$, where $a_{j,0} \define F_{\bx}(q_j)$ and $a_{j,u} \define \frac{\partial F_{\bx}}{z_u}\rvert_{q_j}$, $u \in [\ell]$.
    \item $\{x_i\} \leftarrow {\sf{Extract}}\big(n, i, \{a_j\}_{j \in [k]}, \aux \big)$: The algorithm reconstructs the polynomial $f(y) \define F_{\bx}\big(E(i)+\sum_{s \in [t]}y^s\vs\big)$ and outputs $f(0)=F_{\bx}(E(i))=x_i$. The reconstruction of $f$ is possible because $\deg(f)\leq dt \leq 2k-1$ while $2k$ linear combinations of its coefficients, namely, $\{f(\beta_j),f'(\beta_j)\}_{j \in [k]}$, can be extracted from the answers $\{a_j\}_{j \in [k]}$ and $\aux=\{\vs\}_{s \in [t]}$ as follows. For $j \in [k]$, $f(\beta_j) = F_{\bx}(q_j) = a_{j,0}$, and \vspace{-5pt}
    \[
    f'(\beta_j)\hspace{-2pt} = \hspace{-2pt}\hspace{-2pt}\sum_{u \in [\ell]}\hspace{-2pt} \frac{\partial F}{\partial z_u}\bigg\rvert_{q_j}\hspace{-2pt}\frac{\partial}{\partial y}\bigg(\hspace{-2pt} E(i)_u\hspace{-1pt}+\hspace{-2pt}\sum_{s\in [t]}\hspace{-2pt} y^s\vs_u\hspace{-2pt}\bigg)\hspace{-2pt}\bigg\rvert_{\beta_j}
    \hspace{-5pt}=\hspace{-4pt} \sum_{u \in [\ell]} \hspace{-2pt}a_{j,u}\frac{\partial}{\partial y}\bigg(\hspace{-2pt} E(i)_u+\hspace{-2pt}\sum_{s\in [t]} y^s\vs_u\hspace{-2pt}\bigg)\hspace{-2pt}\bigg\rvert_{\beta_j}\hspace{-3pt}.\vspace{-5pt}
    \]
\end{itemize}
Note that each server generates $\ell\in O(n^{1/d})$ linear combinations of $\bx$. Although in Algorithm~1 we let the LMC verify just one linear combination of $\bx$ for simplicity, in its original form, Lai-Malavolta LMC can verify multiple linear combinations using a single witness. 

\textbf{Bitar-El Rouayheb (BE) Scheme}~(\cite{BitarRouayheb_ITW2018}). 
This scheme works slightly different from the previous ones in that the client retrieves a fixed block of $k-t$ components of $\bx$ instead of a single component. All definitions of a PIR scheme can be generalized to this block form in a straightforward manner. Let $\bx = (x_1,\ldots,x_{(k-t)n})\in
\ff^{(k-t)n}$ and assume the client wants to retrieve the $i$th block $\big(x_{(i-1)(k-t)+1},\ldots,x_{i(k-t)}\big)$ for some $i \in [n]$.
\begin{itemize}
    \item $\big(\{q_j\}_{j\in [k]}, \aux \big) \leftarrow {\sf{QueriesGen}}(n, k, i)$: The algorithm picks a $k\times k$ Vandermonde matrix $\bV = \big(\beta_a^{b-1}\big)_{a,b\in [k]}$, where $\{\beta_s\}_{s\in [t]}$ is a set of $k$ distinct elements in $\bbF$. It also picks $t$ random vectors $\{\vs\}_{s \in [t]}\subseteq \ff^{(k-t)n}$ and set the queries to be the rows of the matrix $\bQ=\bV\bM$ given as follows.
    
    \[
    \bQ = 
    \begin{pmatrix}
    q_1\\
    \hline
    q_2\\
    \hline
    \vdots\\
    \hline
    q_k
    \end{pmatrix}
    \define
    \bV\bM = 
    \bV 
    \begin{pmatrix}
        \vo \\
        \hline
        \vdots \\
        \hline
        \vt \\
        \hline
        \bee_{(i-1)(k-t)+1}\\
        \hline
        \vdots\\
        \hline
        \bee_{i(k-t)}
    \end{pmatrix}.
    \]
    \item $a_j \leftarrow {\sf{AnswerGen}}(\bx, q_j)$: The algorithm outputs $a_j \define q_j \cdot \bx$.
    \item $\{x_i\} \leftarrow {\sf{Extract}}\big(n, i, \{a_j\}_{j \in [k]}, \aux \big)$: The algorithm calculates $\bV^{-1}\bQ\bx = \bM \bx$, which gives $\big(x_{(i-1)(k-t)+1},\ldots,x_{i(k-t)}\big)$.
\end{itemize}
The $t$-privacy is guaranteed because any set of $t$ queries has $t$ random vectors $\{\vs\}_{s \in [t]}$ well mixed (the submatrix of $\bQ$ formed by any $t$ rows and the first $t$ columns is always invertible) and hence appears completely random.

\vspace{-10pt}
\section*{Appendix C: Extra Performance Evaluations of Com-PIR}
\label{app:figures}
\vspace{-5pt}

\begin{figure} [hbt!]
    \centering
        \hspace{-11pt}
        \includegraphics[scale=0.36]{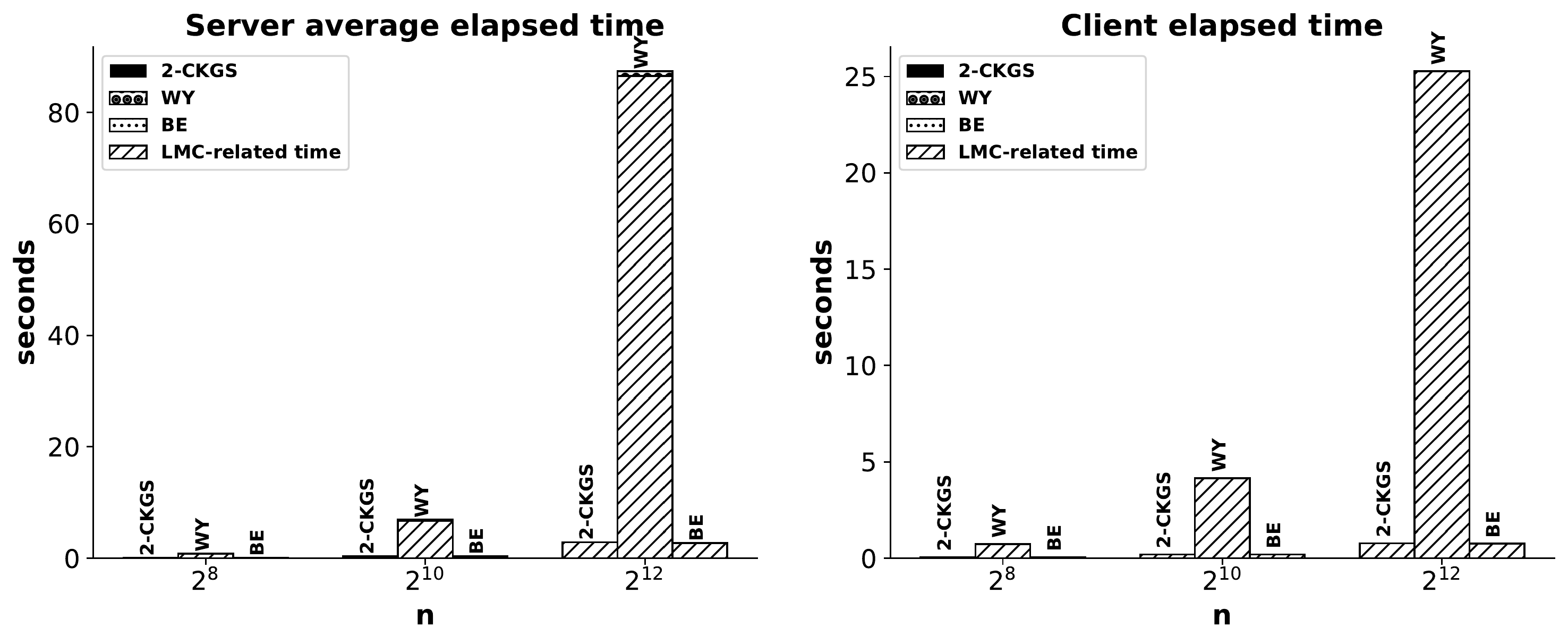}
        \vspace{-15pt}
        \caption{The comparison of the average server and client computation times of LM-CKGS, LM-WY, and LM-BE for $k = 2$, $t = 1$, $m = 2^{10}$, and $n \in \{2^{8},2^{10}, 2^{12}\}$.}
        \label{fig: case2}
\end{figure}

\begin{figure} [hbt!]
    \centering
        \hspace{-11pt}
        \includegraphics[scale=0.36]{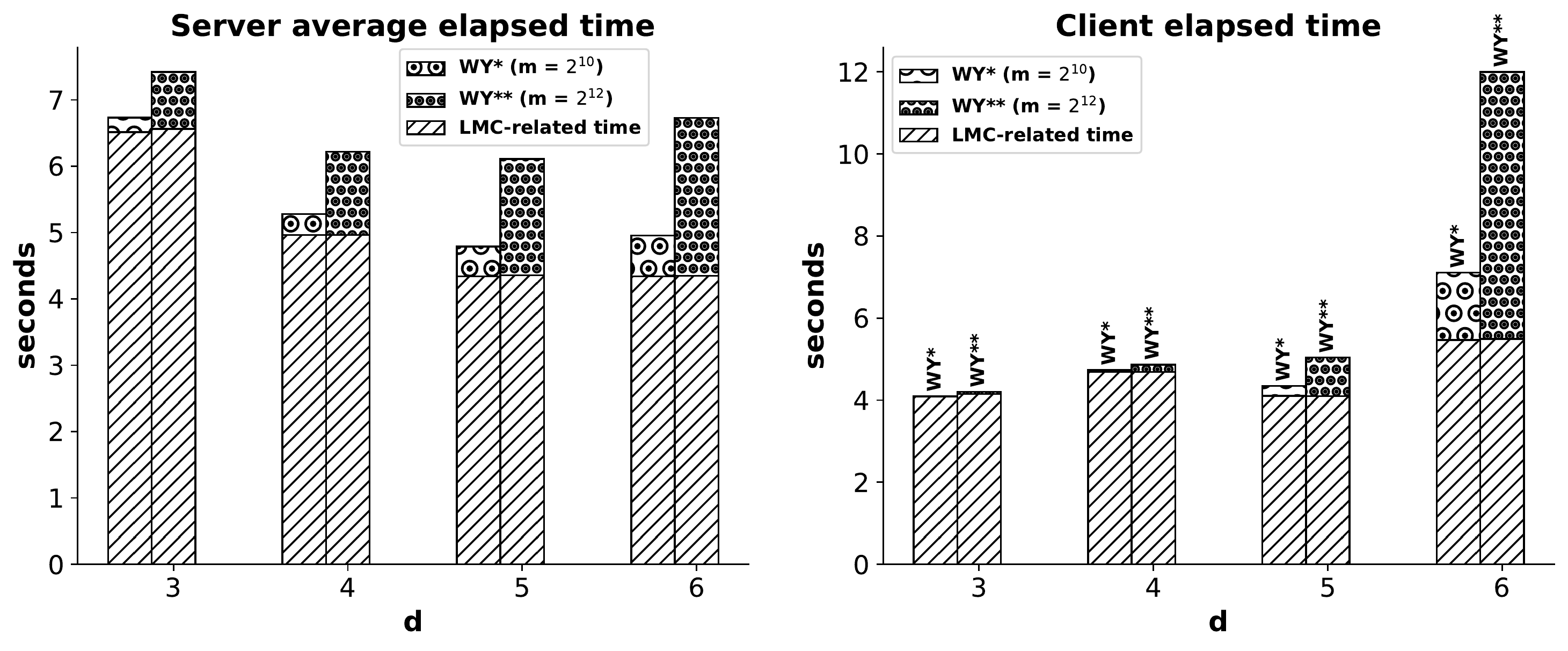}
        \vspace{-15pt}
        \caption{The comparison of the average server and client computation times of LM-WY for $t = 1$, $n = 2^{10}$, $m \in \{2^{10}, 2^{12}\}$, and $(d, k) \in \{(3, 2), (4, 3), (5, 3), (6, 4)\}$.}
        \label{fig: case3}
\end{figure}

In Fig.~\ref{fig: case2}, with a medium $m$ and a growing $n$, the computation time of Com-PIR is dominated by the LMC. Note that LMC on its own scales linearly for the verifier and quadratically in $n$ for the server. More specifically, in our implementation, $O(n)$ elliptic curve operations (expensive) and $O(n^2)$ field operations (cheaper) are required for the server.
Applied on top of a PIR, the LMC-related running time also depends on $k$, $t$, and $\ell\in O(n^{1/d})$. Lai-Malavolta LMC runs reasonably fast on small and medium $n$ (thousands) but slow on larger $n$.
In Fig.~\ref{fig: case3}, we plot the running times of LM-WY as $d$ increases, which means that $k$ increases (to satisfy $2k-1\geq td$ for a fixed $t$) and $\ell$ decreases. While the PIR time increases for both servers and clients, the LMC time for servers (depending on $\ell$) decreases as $d$ grows. For the client, the LMC time fluctuates as it depends on $k\ell$.

\vspace{-15pt}
\vspace{-40pt}

\end{document}